\documentclass[%
 aps,prl,twocolumn,superscriptaddress,floatfix
]{revtex4-2}

\usepackage{graphicx}
\usepackage{dcolumn}
\usepackage{bm}
\usepackage[utf8]{inputenc}
\usepackage[T1]{fontenc}
\usepackage{etoolbox}
\usepackage{newtxtext}
\usepackage[varvw]{newtxmath}

\usepackage{color} 
\makeatletter
\def\@email#1#2{%
 \endgroup
 \patchcmd{\titleblock@produce}
  {\frontmatter@RRAPformat}
  {\frontmatter@RRAPformat{\produce@RRAP{*#1\href{mailto:#2}{#2}}}\frontmatter@RRAPformat}
  {}{}
}%
\makeatother

\begin{document}

\title{Direct observation of the transverse near field of an edge excitation \\ in a fractional quantum Hall state
}

\author{Yunhyeon Jeong}
\affiliation{Department of Physics, Tohoku University, Sendai 980-8578, Japan}

\author{Akinori Kamiyama}
\affiliation{Department of Physics, Tohoku University, Sendai 980-8578, Japan}

\author{John N. Moore}
\affiliation{Department of Physics, Tohoku University, Sendai 980-8578, Japan}

\author{Takaaki Mano}
\affiliation{National Institute for Materials Science, Tsukuba, Ibaraki 305-0047, Japan}

\author{Ken-ichi Sasaki}
\affiliation{NTT Research Center for Theoretical Quantum Information, NTT, Inc., 3-1 Morinosato Wakamiya, Atsugi, Kanagawa 243-0198, Japan}

\author{Yuuki Sugiyama}
\affiliation{Institute for Solid State Physics, The University of Tokyo, Chiba, 277-8581, Japan}

\author{Tokiro Numasawa}
\affiliation{Institute for Solid State Physics, The University of Tokyo, Chiba, 277-8581, Japan}

\author{Masahiro Hotta}
\affiliation{Department of Physics, Tohoku University, Sendai 980-8578, Japan}
\affiliation{Leung Center for Cosmology and Particle Astrophysics, National Taiwan University, Taipei 10617, Taiwan (R.O.C.)}

\author{Go Yusa}
\affiliation{Department of Physics, Tohoku University, Sendai 980-8578, Japan}

\date{\today}

\begin{abstract}

Stroboscopic time-resolved photoluminescence (PL) microscopy and spectroscopy reveal that an electrically launched edge excitation in a $\nu=1/3$ fractional quantum Hall (FQH) state produces an immediate PL response extending more than $30~\mu\mathrm{m}$ into the bulk transverse to the edge when the edge magnetoplasmon (EMP) passes the mesa boundary. The nearly instantaneous nature and downstream-only appearance of this long-range response identify it as the non-radiative, quasi-electrostatic near field.
We also observe a broad delayed response near the mesa boundary that evolves on a much slower time scale and gradually extends into the bulk. The coexistence of the immediate near-field response and this broad delayed response shows that an electrically launched edge excitation cannot be understood solely as a one-dimensional mode propagating along the boundary, but must instead be viewed as a structure extending in both space and time into the surrounding FQH fluid.
\end{abstract}

\maketitle
Fractional quantum Hall (FQH) states \cite{tsui1982two,laughlin1983anomalous} are a paradigmatic realization of strongly correlated topological matter, in which electron-electron interactions give rise to an incompressible bulk with fractionalized quasiparticle excitations \cite{depicciotto1997direct,saminadayar1997observation} and chiral edge degrees of freedom \cite{wen2017colloquium}. The coexistence of a gapped bulk and gapless boundary excitations makes FQH systems a natural platform for studying how a strongly correlated topological phase is reflected in its boundary physics.

Quantum Hall (QH) edge channels provide a versatile platform for coherent electronic interferometry and electron quantum optics \cite{ji2003electronic,feve2007demand,bocquillon2013coherence}, as well as for chiral heat transport \cite{granger2009observation,banerjee2017observed}. These capabilities have motivated broader directions including flying-qubit architectures \cite{assouline2023emission}, quantum energy teleportation \cite{yusa2011quantum,HottaPRA14}, and analog simulations of expanding-universe dynamics \cite{HottaPRD22,Hegde,NambuPRD23,YOSHIMOTO2025130100,sugiyama2025anomalous}.
At the same time, edge dynamics in a FQH system should not be reduced to one-dimensional transport along the boundary. Rather, the edge is the boundary manifestation of an incompressible many-body state \cite{wen1995topological,wen2017colloquium}, and an edge excitation can involve not only chiral charge propagation but also electromagnetic fields and responses of the surrounding QH fluid. Directly resolving this spatially extended structure is therefore essential for understanding edge dynamics as part of the full FQH medium.

\begin{figure}
    \centering
    \includegraphics{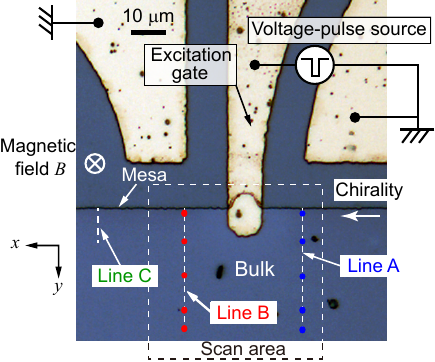}
    \caption{
    Optical microscope image of the measurement device. The excitation gate was driven by a voltage-pulse source  
    (see SM for the details). The dashed rectangle labeled ``Scan area'' corresponds to Fig.~\ref{fig:excitation-gate}. Lines A, B, and C indicate the measurement paths for Figs.~\ref{fig:microPLspectra} and \ref{fig:y-time}. Grounded electrodes are part of coplanar waveguide structures \cite{france2025electrically}.}%
\label{fig:device}
\end{figure}

Among dynamical edge excitations, those launched electrically in QH systems have been widely studied as edge magnetoplasmons (EMPs), collective low-frequency edge modes propagating along the boundary \cite{Ashoori,Ernst,Kamata,hashisaka,matsuuraAPL18,frigerio}. Since an edge excitation involves charge motion, it is naturally accompanied by an electromagnetic field, as is also implicit in early EMP theories \cite{Volkov,Aleiner}. Experimentally, however, EMPs have usually been viewed and probed as one-dimensional charge-density waves propagating along the edge.

Such studies, like QH research more broadly, have relied predominantly on electrical measurements, whereas optical spectroscopy offers capabilities difficult to realize with conventional electrical measurements. Because photoexcited excitons are strongly affected by the surrounding two-dimensional electron system, changes in their optical spectra have served as sensitive probes of many-body correlations, spin states, and electronic compressibility in QH systems and related correlated topological states \cite{yusaPRL01,schuller2003optical,moore2017optically,popert2022optical,cai2023signatures,anderson2024trion}. Optical microscopy also enables spatially resolved measurements and, with ultrafast laser pulses, can follow electrically induced dynamics \cite{kamiyamaPRR,france2025electrically}, with temporal resolution reaching the picosecond range \cite{kamiyamaAPL}.

Here, we use space- and time-resolved PL microscopy and spectroscopy to investigate an electrically launched edge excitation in the $\nu=1/3$ FQH state and observe a long-range transverse optical response. The response extends tens of micrometers into the bulk and remains observable more than $40~\mu\mathrm{m}$ downstream from the excitation gate, consistent with a response arising from the near-field component of an EMP.
Time-resolved measurements further reveal a broad delayed response near the mesa boundary that gradually extends into the bulk. Together with the immediate long-range transverse response, these observations show that an edge excitation cannot be understood solely as a one-dimensional mode propagating along the boundary, but must instead be viewed as a structure extending in both space and time into the surrounding FQH fluid.

The device was fabricated from a wafer containing a 15-nm GaAs/Al$_{0.2}$Ga$_{0.8}$As quantum well (QW) \cite{kamiyamaPRR}, grown on a Si-doped substrate that serves as a back gate for tuning the electron density $n$ in the QW. All measurements reported here were performed in the $\nu = 1/3$ FQH state at $B = 14$~T and a temperature of $40$--$50$~mK. This $\nu = 1/3$ Laughlin state \cite{laughlin1983anomalous} was chosen because it is the simplest Laughlin FQH state, whose edge contains a single chiral charge mode in the idealized low-energy description, and because it provides well-resolved singlet and triplet trion PL signals for local optical probing \cite{yusaPRL01}.
The excitation gate (Fig.~\ref{fig:device}) was driven by a voltage-pulse source; the source configuration and biasing details are given in the Supplemental Material (SM). The excitation gate is a Schottky gate capacitively coupled to the 2DEG; applying a voltage pulse launches an EMP wave packet that propagates leftward along the mesa edge (see End Matter, Appendix A). In all measurements reported here, the rectangular component of the voltage pulse had a duration of approximately $2$~ns.

Laser pulses generated by a mode-locked Ti:sapphire laser were synchronized to the voltage-pulse source. The laser pulse width was $\sim 1$~ps, and the repetition period $T_{\mathrm{rep}}$ was either $\sim 13$~ns or $\sim 52$~ns when a pulse picker was used to reduce the laser repetition rate.
The laser beam was focused onto the sample through an objective lens, yielding a diffraction-limited spot size of $0.77~\mu$m, and the emitted photoluminescence (PL) was collected by the same objective and sent to a spectrometer equipped with a CCD detector. The focal position was controlled using piezoelectric stages. Each spectrum was accumulated over $N_{\mathrm{rep}} = T_{\mathrm{exp}}/T_{\mathrm{rep}}$ repetitions, with a total exposure time of $T_{\mathrm{exp}} = 10$--$20$~s. The relative delay between the laser and voltage pulses was controlled electronically to enable stroboscopic time-resolved measurements. Throughout the manuscript, $t$ denotes the delay time 
for the corresponding measurement configuration. The absolute origin of $t$ is defined separately for each configuration, as described in the SM.

Around $\nu = 1/3$, the PL spectrum contains singlet and triplet trion features associated with different electron-spin configurations \cite{wojs2000charged,yusaPRL01}. Although the detailed line shape in the FQH regime is affected by many-body correlations and is not modeled quantitatively here, we use changes in the trion spectra to probe the local electronic response induced by the edge excitation.

\begin{figure}
    \centering
    \includegraphics[scale=1.0, pagebox=artbox, clip]{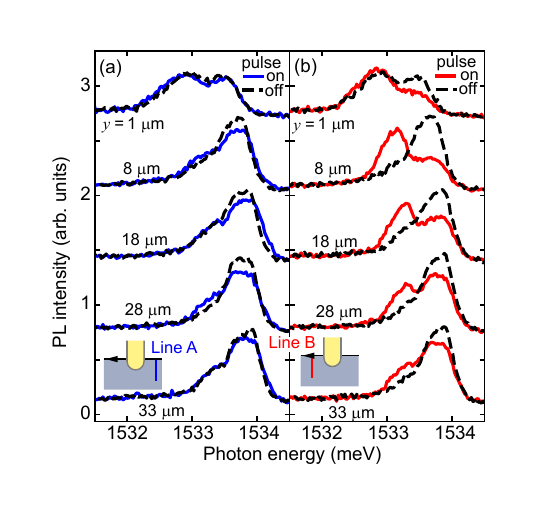}

    \caption{(a) and (b) Micro-PL spectra measured as snapshots at positions along the $y$ direction, perpendicular to the mesa boundary defined by $y=0$: $y=1$, $8$, $18$, $28$, and $33~\mu$m. The delay time is $t=-3.0$~ns for both panels (see SM for more details).
    (a) Spectra measured along Line~A, $16~\mu$m upstream from the center of the excitation gate. (b) Spectra measured along Line~B, $16~\mu$m downstream from the center of the excitation gate. The dotted curves show spectra measured without the excitation pulse. Insets in (a) and (b) schematically indicate the measurement locations (Lines~A and B), as defined in Fig.~\ref{fig:device}. The pulse amplitude is $\pm 100$~mV for both panels. 
    The spectra shown in (b) were measured along Line B under the same voltage-pulse configuration and $T_\mathrm{rep} \sim 13$~ns condition as the time-resolved measurement presented later in Fig.~\ref{fig:y-time}(b).
    }

    \label{fig:microPLspectra}%
\end{figure}

First, to characterize the spatial dependence of the edge-excitation-induced PL response, we recorded micro-PL spectra at five transverse positions along Lines~A and B, as defined in Fig.~\ref{fig:device}.
The PL spectra were recorded at $t = -3.0$~ns. Since the pulse period is $\sim 13$~ns, this timing is equivalent to approximately $10$~ns after the edge excitation reached $y=0$ on Line~B in the preceding cycle.
Along Line~A, located upstream of the excitation gate, the PL spectra remain almost unchanged regardless of whether periodic voltage pulses are applied (blue lines) or not (black dotted lines), as shown in Fig.~\ref{fig:microPLspectra}(a). In contrast, along Line~B, located downstream of the excitation gate, the spectra differ markedly from those along Line~A. Specifically, along Line~B, the application of periodic voltage pulses enhances the low-energy singlet feature and suppresses the high-energy triplet feature (Fig.~\ref{fig:microPLspectra}(b)). This tendency is most pronounced near the mesa boundary at $y\approx 8~\mu$m; at $y\approx 18~\mu$m the singlet intensity still exceeds the triplet intensity, and the singlet feature remains visible even at $28~\mu$m from the mesa boundary.
It is remarkable that signatures of the edge excitation remain visible even at points $28~\mu$m into the bulk and as late as $10$~ns after the excitation reached $y=0$ on Line~B. Since no comparable spectral change is observed upstream, this behavior is unlikely to arise from simple heating.

\begin{figure}
    \includegraphics[scale=1.0, pagebox=artbox, clip]{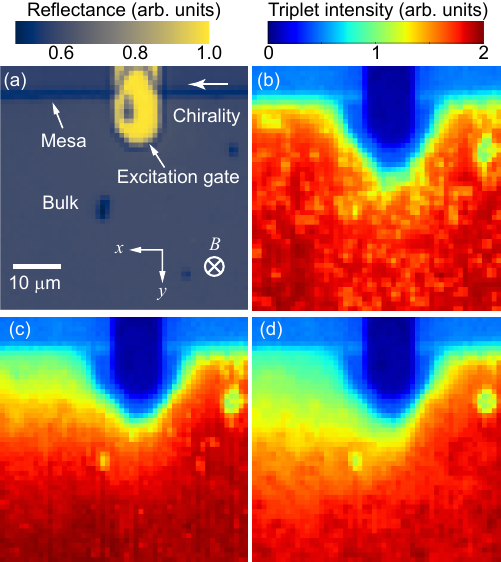}
    \centering
    \caption{
    $51\times 50~\mu\mathrm{m}^2$ spatial maps measured in the region indicated as the scan area in Fig.~\ref{fig:device}. (a) Reflectance map acquired with the same optical setup at $40$--$50$~mK; the yellow region indicates the excitation gate. (b)–(d) Triplet PL intensity maps of the same region measured as snapshots at $t = -3.0$~ns with an exposure time of $10$~s. Here, the triplet PL intensity was obtained by integrating the PL spectrum over $1533.45$--$1534.70$~meV, without spectral fitting. (b) Triplet PL intensity map recorded with no voltage pulse applied to the excitation gate. (c) and (d) Maps obtained with a bipolar voltage pulse 
    applied to the excitation gate. The pulse amplitude is $\pm 50$~mV in panel (c) and $\pm 100$~mV in panel (d). The dark blue region in Figs.~\ref{fig:excitation-gate}(b)--\ref{fig:excitation-gate}(d) corresponds to the region occluded by the excitation gate.}
    \label{fig:excitation-gate}%
\end{figure}

To confirm the spectral modification induced by the edge excitation, we performed similar micro-PL measurements over the scan area indicated by the dashed rectangle in Fig.~\ref{fig:device}, including Lines A and B. Figures~\ref{fig:excitation-gate}(b)--(d) show snapshots of the triplet PL intensity measured at $t=-3.0$~ns, corresponding to a delay of approximately $10$~ns from the time when the edge excitation passed $y=0$ on Line~B.

Without a voltage pulse applied to the excitation gate (Fig.~\ref{fig:excitation-gate}(b)), the PL intensity is nearly symmetric on the upstream and downstream sides of the gate, aside from surface contaminants, which can be identified in the reflectance map (Fig.~\ref{fig:excitation-gate}(a)) and fine spatial fluctuations consistent with a weak disorder-induced potential landscape \cite{eytan1998near,ilani2004microscopic,hayakawa2013real}. Such disorder is commonly attributed to remote ionized donors in modulation-doped heterostructures, for example, Si donors in $\delta$-doping layers separated from the QW \cite{eytan1998near}. The characteristic energy scale of this disorder potential is on the order of $0.1$~meV, as reported in Ref.~\cite{hayakawa2013real}.

When the voltage pulse is applied, the triplet PL intensity on the downstream side is suppressed, and the affected region extends more than $10~\mu$m from the mesa boundary in the $y$ direction (Figs.~\ref{fig:excitation-gate}(c) and \ref{fig:excitation-gate}(d)), whereas no comparable change is observed upstream. Increasing the pulse amplitude from $\pm 50$~mV (Fig.~\ref{fig:excitation-gate}(c)) to $\pm 100$~mV (Fig.~\ref{fig:excitation-gate}(d)) enhances this downstream suppression. At the same time, the contrast associated with the random potential becomes less visible across the image, consistent with heating induced by the voltage pulse.

In this stroboscopic measurement, because the PL spectra are accumulated at $T_{\mathrm{rep}}\sim 13$~ns, phenomena with relaxation times longer than $13$~ns cannot be captured in time. To investigate the origin of the edge-excitation-induced influence over a wide bulk region, we therefore increased $T_{\mathrm{rep}}$ from $\sim 13$~ns to $\sim 52$~ns, i.e., by a factor of four, and examined the possible contribution of longer relaxation processes to the observed decay.

\begin{figure}
    \includegraphics[scale=1.0, pagebox=artbox, clip]{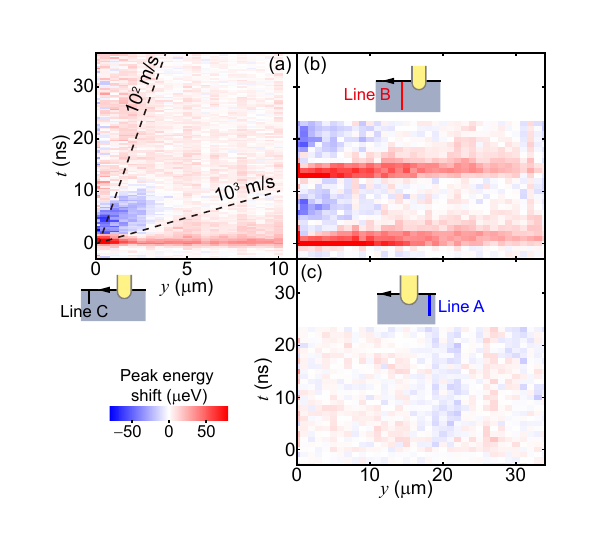}
    \centering
    \caption{Triplet PL peak-energy shift [(a)--(c)], shown as color maps as functions of $y$ and $t$. 
    Panel (a) was measured along Line~C, $\sim 42~\mu$m downstream of the excitation gate, with $T_{\mathrm{rep}} \sim 52$~ns; panel (b) was measured along Line~B, $16~\mu$m downstream, with $T_{\mathrm{rep}} \sim 13$~ns. Panel (c) was measured along Line~A, $16~\mu$m upstream, under the same conditions as panel (b). The inset in each panel schematically indicates the corresponding measurement location (Lines C, B, and A for panels (a), (b), and (c), respectively), as defined in Fig.~\ref{fig:device}.
    The absolute origin of $t$ is defined separately for the Line-C and Line-B measurements; see the SM for details. The voltage-pulse configuration and biasing conditions are described in the SM.
    }
    \label{fig:y-time}%
\end{figure}

The PL spectra were recorded at spatial positions scanned along the $y$ direction on Line C, located $42~\mu\mathrm{m}$ downstream from the center of the excitation gate. 
From these spectra, the triplet peak-energy shift was extracted as a function of $t$ and $y$ and is shown in Fig.~\ref{fig:y-time}(a). The spectral-fitting procedure and the definition of the peak-energy shift are described in the SM. For comparison, the same analysis was applied to data acquired along Line B, with $T_{\mathrm{rep}} \sim 13$~ns [Fig.~\ref{fig:y-time}(b)], and along Line A under the same conditions [Fig.~\ref{fig:y-time}(c)].

The edge excitation is clearly visible in Figs.~\ref{fig:y-time}(a) and \ref{fig:y-time}(b) around $t \sim 0$~ns and $y \sim 0~\mu$m, where the mesa boundary is located. Another notable feature is a triplet peak energy shift extending over nearly the entire $y$ range around $t \sim 0$, appearing as a horizontal red line in both Figs.~\ref{fig:y-time}(a) and (b). In Fig.~\ref{fig:y-time}(b), two such horizontal lines are visible because responses from two periods are captured.  
The associated PL modulation extends more than $30~\mu$m into the bulk [Fig.~~\ref{fig:y-time}], far exceeding microscopic length scales associated with the edge charge distribution, such as 
the magnetic length $l_B=\sqrt{\hbar/eB}=6.9$~nm and the compressible-strip width $a\sim0.1$--$0.3~\mu$m \cite{ilani2004microscopic}.

If the observed response were due to a bulk excitation propagating with a finite transverse velocity, one would expect a delay that increases with $y$, appearing as a slanted feature emerging from $t= 0$~ns and $y = 0~\mu$m in the $y$-$t$ map. Instead, the triplet peak energy shift forms a horizontal feature: it appears at the same timing as the signal from the launched edge excitation near the mesa boundary and extends into the bulk without an observable transverse delay. For a transverse distance of 30~$\mu$m, the electromagnetic propagation time is approximately 0.10~ps in vacuum and 0.36~ps in GaAs, where the latter is estimated using $c/\sqrt{\epsilon_r}$ with $\epsilon_r \simeq 12.9$. Both are far shorter than the time scale resolved in the present measurement. Moreover, upstream Line-A data measured in the same manner as Fig.~\ref{fig:y-time}(b) show no clear time-dependent triplet peak energy shift [Fig.~\ref{fig:y-time}(c)]. This comparison supports the interpretation that the immediate transverse response is associated with the downstream-propagating edge excitation rather than a direct influence of the excitation gate or a spatially uniform optical or electrical artifact. If the signal were caused directly by an electromagnetic wave emitted by the voltage pulse, it should appear both upstream and downstream, whereas it is observed only downstream. We therefore identify this horizontal feature in the $y$-$t$ maps as the transverse near field accompanying the EMP.

Here, “near field” refers to the non-radiative, quasi-electrostatic component of the electromagnetic near field accompanying the EMP charge-density oscillation. To describe the electrostatic kernel of the EMP field, we use the Fredholm-type integral equation in Ref.~\cite{Aleiner} [their Eq.~(6)], whose kernel is given by the modified Bessel function $K_0(|k||y-y_1|)$. The same Coulomb kernel also applies to the charged edge mode at $\nu=1/3$ (see End Matter, Appendix B). Thus, while the charge-density oscillation is concentrated within the boundary strip of width $a$, the electrostatic response associated with a long-wavelength component extends into the bulk over a characteristic transverse length scale of order $\sim 1/|k|$.

The characteristic EMP wavenumber $|k|$ is set by the temporal bandwidth of the launched excitation: the excitation-gate waveform has $\tau \sim 0.5$--$2$~ns (set by the $\sim 2$~ns pulse width and the $\sim 200$~ps rise/fall times), corresponding to frequency components on the order of $f\sim 0.5$--$2$~GHz. Using a representative propagation speed $v\sim 10^{5}$~m/s gives $\lambda \sim v/f \sim 50$--$200~\mu$m and hence $1/|k|=\lambda/2\pi \approx 8$--$32~\mu$m, consistent with the observed long-range modulation of the trion PL.

The EMP is a collective excitation in which charge-density oscillations are inseparably accompanied by electromagnetic fields. Nevertheless, its transverse near-field component has been explored experimentally far less than EMP propagation along the edge. This is largely because the near field is non-radiative, and conventional electrical probes predominantly access signals integrated along the propagation direction, with limited sensitivity to the spatial distribution of the field extending into the bulk \cite{Ashoori,Ernst,Kamata,hashisaka,matsuuraAPL18,frigerio}. In contrast, the trion PL response provides a local optical readout of the 2DEG environment, enabling direct visualization of the long-range transverse influence associated with the EMP near field.

Apart from the immediate transverse near-field response discussed above, a broad delayed response is visible near the mesa boundary in the 52-ns data [Fig.~\ref{fig:y-time}(a)]. This response evolves over several tens of nanoseconds and extends gradually into the bulk. Although this response is broad and lacks a sharp propagating front, its fastest visible spatial evolution corresponds to an apparent rate no greater than $\sim 10^3~\mathrm{m/s}$, substantially slower than the primary bulk magnetoplasmon velocity reported for gate-induced bulk excitations in Ref.~\cite{france2025electrically}.
The present data therefore do not allow us to assign it to a specific propagating mode or relaxation process.

The broad delayed response is clearly visible when $T_{\mathrm{rep}}\sim 52$~ns [Fig.~\ref{fig:y-time}(a)], whereas it is not clearly resolved when $T_{\mathrm{rep}}\sim 13$~ns [Fig.~\ref{fig:y-time}(b)]. This difference is consistent with the slow time scale of the broad delayed response. For the $13$-ns repetition period, the broad delayed response may not fully relax before the next voltage pulse arrives. Under periodic excitation, such incomplete relaxation can produce a cycle-averaged modification of the PL spectrum, rather than a well-resolved time-dependent feature within a single delay window.
In Fig.~\ref{fig:y-time}(a), excluding the region near the edge at $y\simeq 0$ and the blue region associated with the broad delayed response, a faint red region is visible over a wide range of $y$ on the $t>0$ side of the horizontal feature at $t\simeq 0$ (see End Matter, Appendix C). This spatial behavior suggests that the faint red region is associated with the near-field response rather than with the broad delayed response. Since this faint red region may reflect either local relaxation following the near-field response or the finite temporal profile of the near-field response itself, we do not assign it to a separate dynamical mode.

Neither the broad delayed response nor the faint red region can be attributed to trion recombination. The singlet and triplet trion PL decay times are on the order of a few hundred picoseconds \cite{kamiyamaAPL}, much shorter than the time window shown in Fig.~\ref{fig:y-time}.

In light of the above, the PL spectra measured along Line B at $t = -3.0$~ns in Fig.~\ref{fig:microPLspectra}(b) should be interpreted as a cycle-averaged spectral modification, rather than as a snapshot of only the immediate transverse near-field response. This interpretation is supported by the PL spectra shown in Fig.~S2 of the SM (see End Matter, Appendix D). Because this spectral modification does not fully relax before the next voltage pulse is applied to the excitation gate, the spectra in Fig.~\ref{fig:microPLspectra}(b) and Figs.~\ref{fig:excitation-gate}(c) and \ref{fig:excitation-gate}(d) include spectral changes remaining from preceding excitation cycles, rather than showing only the immediate near-field response.

In summary, we investigated electrically launched edge excitations and their accompanying bulk-side responses using space- and time-resolved microscopic PL spectroscopy. Our measurements provide direct experimental evidence that an EMP in the $\nu=1/3$ FQH state produces an immediate transverse response extending more than $30~\mu$m into the bulk, far beyond $l_B$. The response appears without an observable transverse propagation delay and is observed downstream but not upstream. These features identify it as the non-radiative, quasi-electrostatic near-field component accompanying the EMP. In addition, we observed a broad delayed response, indicating that the edge excitation also produces slower dynamics in the surrounding FQH fluid. Together, these observations show that an edge excitation cannot be understood solely as a one-dimensional mode propagating along the boundary, but must instead be viewed as a structure extending in both space and time into the surrounding FQH fluid. More broadly, these findings demonstrate that a dynamical edge mode in a topological electronic system is characterized not only by its propagation along the boundary, but also by its accompanying electromagnetic field and the bulk-side dynamics it induces.

\begin{acknowledgments}
The authors thank T. Takayanagi for fruitful discussions. This work was supported by a Grant-in-Aid for Scientific Research (Grant Nos.~21H05182, 21H05188, and 24H00399) from the Ministry of Education, Culture, Sports, Science, and Technology (MEXT), Japan. T.N. was supported by MEXT KAKENHI Grant Nos.~23K13094 and 24H00944 and by JST PRESTO Grant No.~JPMJPR2359. We are grateful to the YITP workshop YITP-T-25-01 held at YITP, Kyoto University, where part of this work was carried out.
\end{acknowledgments}

\bibliography{biblio}

\begin{thebibliography}{40}%
\makeatletter
\providecommand \@ifxundefined [1]{%
 \@ifx{#1\undefined}
}%
\providecommand \@ifnum [1]{%
 \ifnum #1\expandafter \@firstoftwo
 \else \expandafter \@secondoftwo
 \fi
}%
\providecommand \@ifx [1]{%
 \ifx #1\expandafter \@firstoftwo
 \else \expandafter \@secondoftwo
 \fi
}%
\providecommand \natexlab [1]{#1}%
\providecommand \enquote  [1]{``#1''}%
\providecommand \bibnamefont  [1]{#1}%
\providecommand \bibfnamefont [1]{#1}%
\providecommand \citenamefont [1]{#1}%
\providecommand \href@noop [0]{\@secondoftwo}%
\providecommand \href [0]{\begingroup \@sanitize@url \@href}%
\providecommand \@href[1]{\@@startlink{#1}\@@href}%
\providecommand \@@href[1]{\endgroup#1\@@endlink}%
\providecommand \@sanitize@url [0]{\catcode `\\12\catcode `\$12\catcode `\&12\catcode `\#12\catcode `\^12\catcode `\_12\catcode `\%12\relax}%
\providecommand \@@startlink[1]{}%
\providecommand \@@endlink[0]{}%
\providecommand \url  [0]{\begingroup\@sanitize@url \@url }%
\providecommand \@url [1]{\endgroup\@href {#1}{\urlprefix }}%
\providecommand \urlprefix  [0]{URL }%
\providecommand \Eprint [0]{\href }%
\providecommand \doibase [0]{https://doi.org/}%
\providecommand \selectlanguage [0]{\@gobble}%
\providecommand \bibinfo  [0]{\@secondoftwo}%
\providecommand \bibfield  [0]{\@secondoftwo}%
\providecommand \translation [1]{[#1]}%
\providecommand \BibitemOpen [0]{}%
\providecommand \bibitemStop [0]{}%
\providecommand \bibitemNoStop [0]{.\EOS\space}%
\providecommand \EOS [0]{\spacefactor3000\relax}%
\providecommand \BibitemShut  [1]{\csname bibitem#1\endcsname}%
\let\auto@bib@innerbib\@empty
\bibitem [{\citenamefont {Tsui}\ \emph {et~al.}(1982)\citenamefont {Tsui}, \citenamefont {Stormer},\ and\ \citenamefont {Gossard}}]{tsui1982two}%
  \BibitemOpen
  \bibfield  {author} {\bibinfo {author} {\bibfnamefont {D.~C.}\ \bibnamefont {Tsui}}, \bibinfo {author} {\bibfnamefont {H.~L.}\ \bibnamefont {Stormer}},\ and\ \bibinfo {author} {\bibfnamefont {A.~C.}\ \bibnamefont {Gossard}},\ }\href@noop {} {\bibfield  {journal} {\bibinfo  {journal} {Phys. Rev. Lett.}\ }\textbf {\bibinfo {volume} {48}},\ \bibinfo {pages} {1559} (\bibinfo {year} {1982})}\BibitemShut {NoStop}%
\bibitem [{\citenamefont {Laughlin}(1983)}]{laughlin1983anomalous}%
  \BibitemOpen
  \bibfield  {author} {\bibinfo {author} {\bibfnamefont {R.~B.}\ \bibnamefont {Laughlin}},\ }\href@noop {} {\bibfield  {journal} {\bibinfo  {journal} {Phys. Rev. Lett.}\ }\textbf {\bibinfo {volume} {50}},\ \bibinfo {pages} {1395} (\bibinfo {year} {1983})}\BibitemShut {NoStop}%
\bibitem [{\citenamefont {De-Picciotto}\ \emph {et~al.}(1997)\citenamefont {De-Picciotto}, \citenamefont {Reznikov}, \citenamefont {Heiblum}, \citenamefont {Umansky}, \citenamefont {Bunin},\ and\ \citenamefont {Mahalu}}]{depicciotto1997direct}%
  \BibitemOpen
  \bibfield  {author} {\bibinfo {author} {\bibfnamefont {R.}~\bibnamefont {De-Picciotto}}, \bibinfo {author} {\bibfnamefont {M.}~\bibnamefont {Reznikov}}, \bibinfo {author} {\bibfnamefont {M.}~\bibnamefont {Heiblum}}, \bibinfo {author} {\bibfnamefont {V.}~\bibnamefont {Umansky}}, \bibinfo {author} {\bibfnamefont {G.}~\bibnamefont {Bunin}},\ and\ \bibinfo {author} {\bibfnamefont {D.}~\bibnamefont {Mahalu}},\ }\href@noop {} {\bibfield  {journal} {\bibinfo  {journal} {Nature}\ }\textbf {\bibinfo {volume} {389}},\ \bibinfo {pages} {162} (\bibinfo {year} {1997})}\BibitemShut {NoStop}%
\bibitem [{\citenamefont {Saminadayar}\ \emph {et~al.}(1997)\citenamefont {Saminadayar}, \citenamefont {Glattli}, \citenamefont {Jin},\ and\ \citenamefont {Etienne}}]{saminadayar1997observation}%
  \BibitemOpen
  \bibfield  {author} {\bibinfo {author} {\bibfnamefont {L.}~\bibnamefont {Saminadayar}}, \bibinfo {author} {\bibfnamefont {D.}~\bibnamefont {Glattli}}, \bibinfo {author} {\bibfnamefont {Y.}~\bibnamefont {Jin}},\ and\ \bibinfo {author} {\bibfnamefont {B.}~\bibnamefont {Etienne}},\ }\href@noop {} {\bibfield  {journal} {\bibinfo  {journal} {Phys. Rev. Lett.}\ }\textbf {\bibinfo {volume} {79}},\ \bibinfo {pages} {2526} (\bibinfo {year} {1997})}\BibitemShut {NoStop}%
\bibitem [{\citenamefont {Wen}(2017)}]{wen2017colloquium}%
  \BibitemOpen
  \bibfield  {author} {\bibinfo {author} {\bibfnamefont {X.-G.}\ \bibnamefont {Wen}},\ }\href@noop {} {\bibfield  {journal} {\bibinfo  {journal} {Rev. Mod. Phys.}\ }\textbf {\bibinfo {volume} {89}},\ \bibinfo {pages} {041004} (\bibinfo {year} {2017})}\BibitemShut {NoStop}%
\bibitem [{\citenamefont {Ji}\ \emph {et~al.}(2003)\citenamefont {Ji}, \citenamefont {Chung}, \citenamefont {Sprinzak}, \citenamefont {Heiblum}, \citenamefont {Mahalu},\ and\ \citenamefont {Shtrikman}}]{ji2003electronic}%
  \BibitemOpen
  \bibfield  {author} {\bibinfo {author} {\bibfnamefont {Y.}~\bibnamefont {Ji}}, \bibinfo {author} {\bibfnamefont {Y.}~\bibnamefont {Chung}}, \bibinfo {author} {\bibfnamefont {D.}~\bibnamefont {Sprinzak}}, \bibinfo {author} {\bibfnamefont {M.}~\bibnamefont {Heiblum}}, \bibinfo {author} {\bibfnamefont {D.}~\bibnamefont {Mahalu}},\ and\ \bibinfo {author} {\bibfnamefont {H.}~\bibnamefont {Shtrikman}},\ }\href@noop {} {\bibfield  {journal} {\bibinfo  {journal} {Nature}\ }\textbf {\bibinfo {volume} {422}},\ \bibinfo {pages} {415} (\bibinfo {year} {2003})}\BibitemShut {NoStop}%
\bibitem [{\citenamefont {F{\`e}ve}\ \emph {et~al.}(2007)\citenamefont {F{\`e}ve}, \citenamefont {Mahe}, \citenamefont {Berroir}, \citenamefont {Kontos}, \citenamefont {Placais}, \citenamefont {Glattli}, \citenamefont {Cavanna}, \citenamefont {Etienne},\ and\ \citenamefont {Jin}}]{feve2007demand}%
  \BibitemOpen
  \bibfield  {author} {\bibinfo {author} {\bibfnamefont {G.}~\bibnamefont {F{\`e}ve}}, \bibinfo {author} {\bibfnamefont {A.}~\bibnamefont {Mahe}}, \bibinfo {author} {\bibfnamefont {J.-M.}\ \bibnamefont {Berroir}}, \bibinfo {author} {\bibfnamefont {T.}~\bibnamefont {Kontos}}, \bibinfo {author} {\bibfnamefont {B.}~\bibnamefont {Placais}}, \bibinfo {author} {\bibfnamefont {D.}~\bibnamefont {Glattli}}, \bibinfo {author} {\bibfnamefont {A.}~\bibnamefont {Cavanna}}, \bibinfo {author} {\bibfnamefont {B.}~\bibnamefont {Etienne}},\ and\ \bibinfo {author} {\bibfnamefont {Y.}~\bibnamefont {Jin}},\ }\href@noop {} {\bibfield  {journal} {\bibinfo  {journal} {Science}\ }\textbf {\bibinfo {volume} {316}},\ \bibinfo {pages} {1169} (\bibinfo {year} {2007})}\BibitemShut {NoStop}%
\bibitem [{\citenamefont {Bocquillon}\ \emph {et~al.}(2013)\citenamefont {Bocquillon}, \citenamefont {Freulon}, \citenamefont {Berroir}, \citenamefont {Degiovanni}, \citenamefont {Pla{\c{c}}ais}, \citenamefont {Cavanna}, \citenamefont {Jin},\ and\ \citenamefont {F{\`e}ve}}]{bocquillon2013coherence}%
  \BibitemOpen
  \bibfield  {author} {\bibinfo {author} {\bibfnamefont {E.}~\bibnamefont {Bocquillon}}, \bibinfo {author} {\bibfnamefont {V.}~\bibnamefont {Freulon}}, \bibinfo {author} {\bibfnamefont {J.-M.}\ \bibnamefont {Berroir}}, \bibinfo {author} {\bibfnamefont {P.}~\bibnamefont {Degiovanni}}, \bibinfo {author} {\bibfnamefont {B.}~\bibnamefont {Pla{\c{c}}ais}}, \bibinfo {author} {\bibfnamefont {A.}~\bibnamefont {Cavanna}}, \bibinfo {author} {\bibfnamefont {Y.}~\bibnamefont {Jin}},\ and\ \bibinfo {author} {\bibfnamefont {G.}~\bibnamefont {F{\`e}ve}},\ }\href@noop {} {\bibfield  {journal} {\bibinfo  {journal} {Science}\ }\textbf {\bibinfo {volume} {339}},\ \bibinfo {pages} {1054} (\bibinfo {year} {2013})}\BibitemShut {NoStop}%
\bibitem [{\citenamefont {Granger}\ \emph {et~al.}(2009)\citenamefont {Granger}, \citenamefont {Eisenstein},\ and\ \citenamefont {Reno}}]{granger2009observation}%
  \BibitemOpen
  \bibfield  {author} {\bibinfo {author} {\bibfnamefont {G.}~\bibnamefont {Granger}}, \bibinfo {author} {\bibfnamefont {J.}~\bibnamefont {Eisenstein}},\ and\ \bibinfo {author} {\bibfnamefont {J.}~\bibnamefont {Reno}},\ }\href@noop {} {\bibfield  {journal} {\bibinfo  {journal} {Phys. Rev. Lett.}\ }\textbf {\bibinfo {volume} {102}},\ \bibinfo {pages} {086803} (\bibinfo {year} {2009})}\BibitemShut {NoStop}%
\bibitem [{\citenamefont {Banerjee}\ \emph {et~al.}(2017)\citenamefont {Banerjee}, \citenamefont {Heiblum}, \citenamefont {Rosenblatt}, \citenamefont {Oreg}, \citenamefont {Feldman}, \citenamefont {Stern},\ and\ \citenamefont {Umansky}}]{banerjee2017observed}%
  \BibitemOpen
  \bibfield  {author} {\bibinfo {author} {\bibfnamefont {M.}~\bibnamefont {Banerjee}}, \bibinfo {author} {\bibfnamefont {M.}~\bibnamefont {Heiblum}}, \bibinfo {author} {\bibfnamefont {A.}~\bibnamefont {Rosenblatt}}, \bibinfo {author} {\bibfnamefont {Y.}~\bibnamefont {Oreg}}, \bibinfo {author} {\bibfnamefont {D.~E.}\ \bibnamefont {Feldman}}, \bibinfo {author} {\bibfnamefont {A.}~\bibnamefont {Stern}},\ and\ \bibinfo {author} {\bibfnamefont {V.}~\bibnamefont {Umansky}},\ }\href@noop {} {\bibfield  {journal} {\bibinfo  {journal} {Nature}\ }\textbf {\bibinfo {volume} {545}},\ \bibinfo {pages} {75} (\bibinfo {year} {2017})}\BibitemShut {NoStop}%
\bibitem [{\citenamefont {Assouline}\ \emph {et~al.}(2023)\citenamefont {Assouline}, \citenamefont {Pugliese}, \citenamefont {Chakraborti}, \citenamefont {Lee}, \citenamefont {Bernabeu}, \citenamefont {Jo}, \citenamefont {Watanabe}, \citenamefont {Taniguchi}, \citenamefont {Glattli}, \citenamefont {Kumada} \emph {et~al.}}]{assouline2023emission}%
  \BibitemOpen
  \bibfield  {author} {\bibinfo {author} {\bibfnamefont {A.}~\bibnamefont {Assouline}}, \bibinfo {author} {\bibfnamefont {L.}~\bibnamefont {Pugliese}}, \bibinfo {author} {\bibfnamefont {H.}~\bibnamefont {Chakraborti}}, \bibinfo {author} {\bibfnamefont {S.}~\bibnamefont {Lee}}, \bibinfo {author} {\bibfnamefont {L.}~\bibnamefont {Bernabeu}}, \bibinfo {author} {\bibfnamefont {M.}~\bibnamefont {Jo}}, \bibinfo {author} {\bibfnamefont {K.}~\bibnamefont {Watanabe}}, \bibinfo {author} {\bibfnamefont {T.}~\bibnamefont {Taniguchi}}, \bibinfo {author} {\bibfnamefont {D.}~\bibnamefont {Glattli}}, \bibinfo {author} {\bibfnamefont {N.}~\bibnamefont {Kumada}}, \emph {et~al.},\ }\href@noop {} {\bibfield  {journal} {\bibinfo  {journal} {Science}\ }\textbf {\bibinfo {volume} {382}},\ \bibinfo {pages} {1260} (\bibinfo {year} {2023})}\BibitemShut {NoStop}%
\bibitem [{\citenamefont {Yusa}\ \emph {et~al.}(2011)\citenamefont {Yusa}, \citenamefont {Izumida},\ and\ \citenamefont {Hotta}}]{yusa2011quantum}%
  \BibitemOpen
  \bibfield  {author} {\bibinfo {author} {\bibfnamefont {G.}~\bibnamefont {Yusa}}, \bibinfo {author} {\bibfnamefont {W.}~\bibnamefont {Izumida}},\ and\ \bibinfo {author} {\bibfnamefont {M.}~\bibnamefont {Hotta}},\ }\href@noop {} {\bibfield  {journal} {\bibinfo  {journal} {Phys. Rev. A}\ }\textbf {\bibinfo {volume} {84}},\ \bibinfo {pages} {032336} (\bibinfo {year} {2011})}\BibitemShut {NoStop}%
\bibitem [{\citenamefont {Hotta}\ \emph {et~al.}(2014)\citenamefont {Hotta}, \citenamefont {Matsumoto},\ and\ \citenamefont {Yusa}}]{HottaPRA14}%
  \BibitemOpen
  \bibfield  {author} {\bibinfo {author} {\bibfnamefont {M.}~\bibnamefont {Hotta}}, \bibinfo {author} {\bibfnamefont {J.}~\bibnamefont {Matsumoto}},\ and\ \bibinfo {author} {\bibfnamefont {G.}~\bibnamefont {Yusa}},\ }\href {https://doi.org/10.1103/PhysRevA.89.012311} {\bibfield  {journal} {\bibinfo  {journal} {Phys. Rev. A}\ }\textbf {\bibinfo {volume} {89}},\ \bibinfo {pages} {012311} (\bibinfo {year} {2014})}\BibitemShut {NoStop}%
\bibitem [{\citenamefont {Hotta}\ \emph {et~al.}(2022)\citenamefont {Hotta}, \citenamefont {Nambu}, \citenamefont {Sugiyama}, \citenamefont {Yamamoto},\ and\ \citenamefont {Yusa}}]{HottaPRD22}%
  \BibitemOpen
  \bibfield  {author} {\bibinfo {author} {\bibfnamefont {M.}~\bibnamefont {Hotta}}, \bibinfo {author} {\bibfnamefont {Y.}~\bibnamefont {Nambu}}, \bibinfo {author} {\bibfnamefont {Y.}~\bibnamefont {Sugiyama}}, \bibinfo {author} {\bibfnamefont {K.}~\bibnamefont {Yamamoto}},\ and\ \bibinfo {author} {\bibfnamefont {G.}~\bibnamefont {Yusa}},\ }\href {https://doi.org/10.1103/PhysRevD.105.105009} {\bibfield  {journal} {\bibinfo  {journal} {Phys. Rev. D}\ }\textbf {\bibinfo {volume} {105}},\ \bibinfo {pages} {105009} (\bibinfo {year} {2022})}\BibitemShut {NoStop}%
\bibitem [{\citenamefont {Hegde}\ \emph {et~al.}(2019)\citenamefont {Hegde}, \citenamefont {Subramanyan}, \citenamefont {Bradlyn},\ and\ \citenamefont {Vishveshwara}}]{Hegde}%
  \BibitemOpen
  \bibfield  {author} {\bibinfo {author} {\bibfnamefont {S.~S.}\ \bibnamefont {Hegde}}, \bibinfo {author} {\bibfnamefont {V.}~\bibnamefont {Subramanyan}}, \bibinfo {author} {\bibfnamefont {B.}~\bibnamefont {Bradlyn}},\ and\ \bibinfo {author} {\bibfnamefont {S.}~\bibnamefont {Vishveshwara}},\ }\href {https://doi.org/10.1103/PhysRevLett.123.156802} {\bibfield  {journal} {\bibinfo  {journal} {Phys. Rev. Lett.}\ }\textbf {\bibinfo {volume} {123}},\ \bibinfo {pages} {156802} (\bibinfo {year} {2019})}\BibitemShut {NoStop}%
\bibitem [{\citenamefont {Nambu}\ and\ \citenamefont {Hotta}(2023)}]{NambuPRD23}%
  \BibitemOpen
  \bibfield  {author} {\bibinfo {author} {\bibfnamefont {Y.}~\bibnamefont {Nambu}}\ and\ \bibinfo {author} {\bibfnamefont {M.}~\bibnamefont {Hotta}},\ }\href {https://doi.org/10.1103/PhysRevD.107.085002} {\bibfield  {journal} {\bibinfo  {journal} {Phys. Rev. D}\ }\textbf {\bibinfo {volume} {107}},\ \bibinfo {pages} {085002} (\bibinfo {year} {2023})}\BibitemShut {NoStop}%
\bibitem [{\citenamefont {Yoshimoto}\ and\ \citenamefont {Nambu}(2025)}]{YOSHIMOTO2025130100}%
  \BibitemOpen
  \bibfield  {author} {\bibinfo {author} {\bibfnamefont {R.}~\bibnamefont {Yoshimoto}}\ and\ \bibinfo {author} {\bibfnamefont {Y.}~\bibnamefont {Nambu}},\ }\href {https://doi.org/https://doi.org/10.1016/j.physleta.2024.130100} {\bibfield  {journal} {\bibinfo  {journal} {Phys. Lett. A}\ }\textbf {\bibinfo {volume} {529}},\ \bibinfo {pages} {130100} (\bibinfo {year} {2025})}\BibitemShut {NoStop}%
\bibitem [{\citenamefont {Sugiyama}\ and\ \citenamefont {Numasawa}(2025)}]{sugiyama2025anomalous}%
  \BibitemOpen
  \bibfield  {author} {\bibinfo {author} {\bibfnamefont {Y.}~\bibnamefont {Sugiyama}}\ and\ \bibinfo {author} {\bibfnamefont {T.}~\bibnamefont {Numasawa}},\ }\href@noop {} {\bibfield  {journal} {\bibinfo  {journal} {arXiv:2506.20338}\ } (\bibinfo {year} {2025})}\BibitemShut {NoStop}%
\bibitem [{\citenamefont {Wen}(1995)}]{wen1995topological}%
  \BibitemOpen
  \bibfield  {author} {\bibinfo {author} {\bibfnamefont {X.-G.}\ \bibnamefont {Wen}},\ }\href@noop {} {\bibfield  {journal} {\bibinfo  {journal} {Adv. Phys.}\ }\textbf {\bibinfo {volume} {44}},\ \bibinfo {pages} {405} (\bibinfo {year} {1995})}\BibitemShut {NoStop}%
\bibitem [{\citenamefont {France}\ \emph {et~al.}(2025)\citenamefont {France}, \citenamefont {Jeong}, \citenamefont {Kamiyama}, \citenamefont {Mano}, \citenamefont {Sasaki}, \citenamefont {Hotta},\ and\ \citenamefont {Yusa}}]{france2025electrically}%
  \BibitemOpen
  \bibfield  {author} {\bibinfo {author} {\bibfnamefont {Q.}~\bibnamefont {France}}, \bibinfo {author} {\bibfnamefont {Y.}~\bibnamefont {Jeong}}, \bibinfo {author} {\bibfnamefont {A.}~\bibnamefont {Kamiyama}}, \bibinfo {author} {\bibfnamefont {T.}~\bibnamefont {Mano}}, \bibinfo {author} {\bibfnamefont {K.-i.}\ \bibnamefont {Sasaki}}, \bibinfo {author} {\bibfnamefont {M.}~\bibnamefont {Hotta}},\ and\ \bibinfo {author} {\bibfnamefont {G.}~\bibnamefont {Yusa}},\ }\href@noop {} {\bibfield  {journal} {\bibinfo  {journal} {Phys. Rev. Lett.}\ }\textbf {\bibinfo {volume} {135}},\ \bibinfo {pages} {066203} (\bibinfo {year} {2025})}\BibitemShut {NoStop}%
\bibitem [{\citenamefont {Ashoori}\ \emph {et~al.}(1992)\citenamefont {Ashoori}, \citenamefont {Stormer}, \citenamefont {Pfeiffer}, \citenamefont {Baldwin},\ and\ \citenamefont {West}}]{Ashoori}%
  \BibitemOpen
  \bibfield  {author} {\bibinfo {author} {\bibfnamefont {R.~C.}\ \bibnamefont {Ashoori}}, \bibinfo {author} {\bibfnamefont {H.~L.}\ \bibnamefont {Stormer}}, \bibinfo {author} {\bibfnamefont {L.~N.}\ \bibnamefont {Pfeiffer}}, \bibinfo {author} {\bibfnamefont {K.~W.}\ \bibnamefont {Baldwin}},\ and\ \bibinfo {author} {\bibfnamefont {K.}~\bibnamefont {West}},\ }\href {https://doi.org/10.1103/PhysRevB.45.3894} {\bibfield  {journal} {\bibinfo  {journal} {Phys. Rev. B}\ }\textbf {\bibinfo {volume} {45}},\ \bibinfo {pages} {3894} (\bibinfo {year} {1992})}\BibitemShut {NoStop}%
\bibitem [{\citenamefont {Ernst}\ \emph {et~al.}(1996)\citenamefont {Ernst}, \citenamefont {Haug}, \citenamefont {Kuhl}, \citenamefont {von Klitzing},\ and\ \citenamefont {Eberl}}]{Ernst}%
  \BibitemOpen
  \bibfield  {author} {\bibinfo {author} {\bibfnamefont {G.}~\bibnamefont {Ernst}}, \bibinfo {author} {\bibfnamefont {R.~J.}\ \bibnamefont {Haug}}, \bibinfo {author} {\bibfnamefont {J.}~\bibnamefont {Kuhl}}, \bibinfo {author} {\bibfnamefont {K.}~\bibnamefont {von Klitzing}},\ and\ \bibinfo {author} {\bibfnamefont {K.}~\bibnamefont {Eberl}},\ }\href {https://doi.org/10.1103/PhysRevLett.77.4245} {\bibfield  {journal} {\bibinfo  {journal} {Phys. Rev. Lett.}\ }\textbf {\bibinfo {volume} {77}},\ \bibinfo {pages} {4245} (\bibinfo {year} {1996})}\BibitemShut {NoStop}%
\bibitem [{\citenamefont {Kamata}\ \emph {et~al.}(2010)\citenamefont {Kamata}, \citenamefont {Ota}, \citenamefont {Muraki},\ and\ \citenamefont {Fujisawa}}]{Kamata}%
  \BibitemOpen
  \bibfield  {author} {\bibinfo {author} {\bibfnamefont {H.}~\bibnamefont {Kamata}}, \bibinfo {author} {\bibfnamefont {T.}~\bibnamefont {Ota}}, \bibinfo {author} {\bibfnamefont {K.}~\bibnamefont {Muraki}},\ and\ \bibinfo {author} {\bibfnamefont {T.}~\bibnamefont {Fujisawa}},\ }\href {https://doi.org/10.1103/PhysRevB.81.085329} {\bibfield  {journal} {\bibinfo  {journal} {Phys. Rev. B}\ }\textbf {\bibinfo {volume} {81}},\ \bibinfo {pages} {085329} (\bibinfo {year} {2010})}\BibitemShut {NoStop}%
\bibitem [{\citenamefont {Hashisaka}\ \emph {et~al.}(2017)\citenamefont {Hashisaka}, \citenamefont {Hiyama}, \citenamefont {Akiho}, \citenamefont {Muraki},\ and\ \citenamefont {Fujisawa}}]{hashisaka}%
  \BibitemOpen
  \bibfield  {author} {\bibinfo {author} {\bibfnamefont {M.}~\bibnamefont {Hashisaka}}, \bibinfo {author} {\bibfnamefont {N.}~\bibnamefont {Hiyama}}, \bibinfo {author} {\bibfnamefont {T.}~\bibnamefont {Akiho}}, \bibinfo {author} {\bibfnamefont {K.}~\bibnamefont {Muraki}},\ and\ \bibinfo {author} {\bibfnamefont {T.}~\bibnamefont {Fujisawa}},\ }\href {https://doi.org/10.1038/nphys4062} {\bibfield  {journal} {\bibinfo  {journal} {Nature Phys.}\ }\textbf {\bibinfo {volume} {13}},\ \bibinfo {pages} {559} (\bibinfo {year} {2017})}\BibitemShut {NoStop}%
\bibitem [{\citenamefont {Matsuura}\ \emph {et~al.}(2018)\citenamefont {Matsuura}, \citenamefont {Mano}, \citenamefont {Noda}, \citenamefont {Shibata}, \citenamefont {Hotta},\ and\ \citenamefont {Yusa}}]{matsuuraAPL18}%
  \BibitemOpen
  \bibfield  {author} {\bibinfo {author} {\bibfnamefont {M.}~\bibnamefont {Matsuura}}, \bibinfo {author} {\bibfnamefont {T.}~\bibnamefont {Mano}}, \bibinfo {author} {\bibfnamefont {T.}~\bibnamefont {Noda}}, \bibinfo {author} {\bibfnamefont {N.}~\bibnamefont {Shibata}}, \bibinfo {author} {\bibfnamefont {M.}~\bibnamefont {Hotta}},\ and\ \bibinfo {author} {\bibfnamefont {G.}~\bibnamefont {Yusa}},\ }\href@noop {} {\bibfield  {journal} {\bibinfo  {journal} {Appl. Phys. Lett.}\ }\textbf {\bibinfo {volume} {112}},\ \bibinfo {pages} {063104} (\bibinfo {year} {2018})}\BibitemShut {NoStop}%
\bibitem [{\citenamefont {Frigerio}\ \emph {et~al.}(2024)\citenamefont {Frigerio}, \citenamefont {Rebora}, \citenamefont {Ruelle}, \citenamefont {Souquet-Basi{\`e}ge}, \citenamefont {Jin}, \citenamefont {Gennser}, \citenamefont {Cavanna}, \citenamefont {Pla{\c{c}}ais}, \citenamefont {Baudin}, \citenamefont {Berroir}, \citenamefont {Safi}, \citenamefont {Degiovanni}, \citenamefont {F{\`e}ve},\ and\ \citenamefont {M{\'e}nard}}]{frigerio}%
  \BibitemOpen
  \bibfield  {author} {\bibinfo {author} {\bibfnamefont {E.}~\bibnamefont {Frigerio}}, \bibinfo {author} {\bibfnamefont {G.}~\bibnamefont {Rebora}}, \bibinfo {author} {\bibfnamefont {M.}~\bibnamefont {Ruelle}}, \bibinfo {author} {\bibfnamefont {H.}~\bibnamefont {Souquet-Basi{\`e}ge}}, \bibinfo {author} {\bibfnamefont {Y.}~\bibnamefont {Jin}}, \bibinfo {author} {\bibfnamefont {U.}~\bibnamefont {Gennser}}, \bibinfo {author} {\bibfnamefont {A.}~\bibnamefont {Cavanna}}, \bibinfo {author} {\bibfnamefont {B.}~\bibnamefont {Pla{\c{c}}ais}}, \bibinfo {author} {\bibfnamefont {E.}~\bibnamefont {Baudin}}, \bibinfo {author} {\bibfnamefont {J.-M.}\ \bibnamefont {Berroir}}, \bibinfo {author} {\bibfnamefont {I.}~\bibnamefont {Safi}}, \bibinfo {author} {\bibfnamefont {P.}~\bibnamefont {Degiovanni}}, \bibinfo {author} {\bibfnamefont {G.}~\bibnamefont {F{\`e}ve}},\ and\ \bibinfo {author} {\bibfnamefont {G.~C.}\ \bibnamefont {M{\'e}nard}},\ }\href {https://doi.org/10.1038/s42005-024-01803-6} {\bibfield  {journal} {\bibinfo
  {journal} {Commun. Phys.}\ }\textbf {\bibinfo {volume} {7}},\ \bibinfo {pages} {314} (\bibinfo {year} {2024})}\BibitemShut {NoStop}%
\bibitem [{\citenamefont {Volkov}\ and\ \citenamefont {Mikhailov}(1988)}]{Volkov}%
  \BibitemOpen
  \bibfield  {author} {\bibinfo {author} {\bibfnamefont {V.~A.}\ \bibnamefont {Volkov}}\ and\ \bibinfo {author} {\bibfnamefont {S.~A.}\ \bibnamefont {Mikhailov}},\ }\href@noop {} {\bibfield  {journal} {\bibinfo  {journal} {Sov. Phys. JETP}\ }\textbf {\bibinfo {volume} {67}},\ \bibinfo {pages} {1639} (\bibinfo {year} {1988})}\BibitemShut {NoStop}%
\bibitem [{\citenamefont {Aleiner}\ and\ \citenamefont {Glazman}(1994)}]{Aleiner}%
  \BibitemOpen
  \bibfield  {author} {\bibinfo {author} {\bibfnamefont {I.~L.}\ \bibnamefont {Aleiner}}\ and\ \bibinfo {author} {\bibfnamefont {L.~I.}\ \bibnamefont {Glazman}},\ }\href {https://doi.org/10.1103/PhysRevLett.72.2935} {\bibfield  {journal} {\bibinfo  {journal} {Phys. Rev. Lett.}\ }\textbf {\bibinfo {volume} {72}},\ \bibinfo {pages} {2935} (\bibinfo {year} {1994})}\BibitemShut {NoStop}%
\bibitem [{\citenamefont {Yusa}\ \emph {et~al.}(2001)\citenamefont {Yusa}, \citenamefont {Shtrikman},\ and\ \citenamefont {Bar-Joseph}}]{yusaPRL01}%
  \BibitemOpen
  \bibfield  {author} {\bibinfo {author} {\bibfnamefont {G.}~\bibnamefont {Yusa}}, \bibinfo {author} {\bibfnamefont {H.}~\bibnamefont {Shtrikman}},\ and\ \bibinfo {author} {\bibfnamefont {I.}~\bibnamefont {Bar-Joseph}},\ }\href@noop {} {\bibfield  {journal} {\bibinfo  {journal} {Phys. Rev. Lett.}\ }\textbf {\bibinfo {volume} {87}},\ \bibinfo {pages} {216402} (\bibinfo {year} {2001})}\BibitemShut {NoStop}%
\bibitem [{\citenamefont {Sch{\"u}ller}\ \emph {et~al.}(2003)\citenamefont {Sch{\"u}ller}, \citenamefont {Broocks}, \citenamefont {Schr{\"o}ter}, \citenamefont {Heyn}, \citenamefont {Heitmann}, \citenamefont {Bichler}, \citenamefont {Wegscheider}, \citenamefont {Chakraborty},\ and\ \citenamefont {Apalkov}}]{schuller2003optical}%
  \BibitemOpen
  \bibfield  {author} {\bibinfo {author} {\bibfnamefont {C.}~\bibnamefont {Sch{\"u}ller}}, \bibinfo {author} {\bibfnamefont {K.-B.}\ \bibnamefont {Broocks}}, \bibinfo {author} {\bibfnamefont {P.}~\bibnamefont {Schr{\"o}ter}}, \bibinfo {author} {\bibfnamefont {C.}~\bibnamefont {Heyn}}, \bibinfo {author} {\bibfnamefont {D.}~\bibnamefont {Heitmann}}, \bibinfo {author} {\bibfnamefont {M.}~\bibnamefont {Bichler}}, \bibinfo {author} {\bibfnamefont {W.}~\bibnamefont {Wegscheider}}, \bibinfo {author} {\bibfnamefont {T.}~\bibnamefont {Chakraborty}},\ and\ \bibinfo {author} {\bibfnamefont {V.}~\bibnamefont {Apalkov}},\ }\href@noop {} {\bibfield  {journal} {\bibinfo  {journal} {Phys. Rev. Lett.}\ }\textbf {\bibinfo {volume} {91}},\ \bibinfo {pages} {116403} (\bibinfo {year} {2003})}\BibitemShut {NoStop}%
\bibitem [{\citenamefont {Moore}\ \emph {et~al.}(2017)\citenamefont {Moore}, \citenamefont {Hayakawa}, \citenamefont {Mano}, \citenamefont {Noda},\ and\ \citenamefont {Yusa}}]{moore2017optically}%
  \BibitemOpen
  \bibfield  {author} {\bibinfo {author} {\bibfnamefont {J.~N.}\ \bibnamefont {Moore}}, \bibinfo {author} {\bibfnamefont {J.}~\bibnamefont {Hayakawa}}, \bibinfo {author} {\bibfnamefont {T.}~\bibnamefont {Mano}}, \bibinfo {author} {\bibfnamefont {T.}~\bibnamefont {Noda}},\ and\ \bibinfo {author} {\bibfnamefont {G.}~\bibnamefont {Yusa}},\ }\href@noop {} {\bibfield  {journal} {\bibinfo  {journal} {Phys. Rev. Lett.}\ }\textbf {\bibinfo {volume} {118}},\ \bibinfo {pages} {076802} (\bibinfo {year} {2017})}\BibitemShut {NoStop}%
\bibitem [{\citenamefont {Popert}\ \emph {et~al.}(2022)\citenamefont {Popert}, \citenamefont {Shimazaki}, \citenamefont {Kroner}, \citenamefont {Watanabe}, \citenamefont {Taniguchi}, \citenamefont {Imamoglu},\ and\ \citenamefont {Smolenski}}]{popert2022optical}%
  \BibitemOpen
  \bibfield  {author} {\bibinfo {author} {\bibfnamefont {A.}~\bibnamefont {Popert}}, \bibinfo {author} {\bibfnamefont {Y.}~\bibnamefont {Shimazaki}}, \bibinfo {author} {\bibfnamefont {M.}~\bibnamefont {Kroner}}, \bibinfo {author} {\bibfnamefont {K.}~\bibnamefont {Watanabe}}, \bibinfo {author} {\bibfnamefont {T.}~\bibnamefont {Taniguchi}}, \bibinfo {author} {\bibfnamefont {A.}~\bibnamefont {Imamoglu}},\ and\ \bibinfo {author} {\bibfnamefont {T.}~\bibnamefont {Smolenski}},\ }\href@noop {} {\bibfield  {journal} {\bibinfo  {journal} {Nano Lett.}\ }\textbf {\bibinfo {volume} {22}},\ \bibinfo {pages} {7363} (\bibinfo {year} {2022})}\BibitemShut {NoStop}%
\bibitem [{\citenamefont {Cai}\ \emph {et~al.}(2023)\citenamefont {Cai}, \citenamefont {Anderson}, \citenamefont {Wang}, \citenamefont {Zhang}, \citenamefont {Liu}, \citenamefont {Holtzmann}, \citenamefont {Zhang}, \citenamefont {Fan}, \citenamefont {Taniguchi}, \citenamefont {Watanabe} \emph {et~al.}}]{cai2023signatures}%
  \BibitemOpen
  \bibfield  {author} {\bibinfo {author} {\bibfnamefont {J.}~\bibnamefont {Cai}}, \bibinfo {author} {\bibfnamefont {E.}~\bibnamefont {Anderson}}, \bibinfo {author} {\bibfnamefont {C.}~\bibnamefont {Wang}}, \bibinfo {author} {\bibfnamefont {X.}~\bibnamefont {Zhang}}, \bibinfo {author} {\bibfnamefont {X.}~\bibnamefont {Liu}}, \bibinfo {author} {\bibfnamefont {W.}~\bibnamefont {Holtzmann}}, \bibinfo {author} {\bibfnamefont {Y.}~\bibnamefont {Zhang}}, \bibinfo {author} {\bibfnamefont {F.}~\bibnamefont {Fan}}, \bibinfo {author} {\bibfnamefont {T.}~\bibnamefont {Taniguchi}}, \bibinfo {author} {\bibfnamefont {K.}~\bibnamefont {Watanabe}}, \emph {et~al.},\ }\href@noop {} {\bibfield  {journal} {\bibinfo  {journal} {Nature}\ }\textbf {\bibinfo {volume} {622}},\ \bibinfo {pages} {63} (\bibinfo {year} {2023})}\BibitemShut {NoStop}%
\bibitem [{\citenamefont {Anderson}\ \emph {et~al.}(2024)\citenamefont {Anderson}, \citenamefont {Cai}, \citenamefont {Reddy}, \citenamefont {Park}, \citenamefont {Holtzmann}, \citenamefont {Davis}, \citenamefont {Taniguchi}, \citenamefont {Watanabe}, \citenamefont {Smolenski}, \citenamefont {Imamo{\u{g}}lu} \emph {et~al.}}]{anderson2024trion}%
  \BibitemOpen
  \bibfield  {author} {\bibinfo {author} {\bibfnamefont {E.}~\bibnamefont {Anderson}}, \bibinfo {author} {\bibfnamefont {J.}~\bibnamefont {Cai}}, \bibinfo {author} {\bibfnamefont {A.~P.}\ \bibnamefont {Reddy}}, \bibinfo {author} {\bibfnamefont {H.}~\bibnamefont {Park}}, \bibinfo {author} {\bibfnamefont {W.}~\bibnamefont {Holtzmann}}, \bibinfo {author} {\bibfnamefont {K.}~\bibnamefont {Davis}}, \bibinfo {author} {\bibfnamefont {T.}~\bibnamefont {Taniguchi}}, \bibinfo {author} {\bibfnamefont {K.}~\bibnamefont {Watanabe}}, \bibinfo {author} {\bibfnamefont {T.}~\bibnamefont {Smolenski}}, \bibinfo {author} {\bibfnamefont {A.}~\bibnamefont {Imamo{\u{g}}lu}}, \emph {et~al.},\ }\href@noop {} {\bibfield  {journal} {\bibinfo  {journal} {Nature}\ }\textbf {\bibinfo {volume} {635}},\ \bibinfo {pages} {590} (\bibinfo {year} {2024})}\BibitemShut {NoStop}%
\bibitem [{\citenamefont {Kamiyama}\ \emph {et~al.}(2022)\citenamefont {Kamiyama}, \citenamefont {Matsuura}, \citenamefont {Moore}, \citenamefont {Mano}, \citenamefont {Shibata},\ and\ \citenamefont {Yusa}}]{kamiyamaPRR}%
  \BibitemOpen
  \bibfield  {author} {\bibinfo {author} {\bibfnamefont {A.}~\bibnamefont {Kamiyama}}, \bibinfo {author} {\bibfnamefont {M.}~\bibnamefont {Matsuura}}, \bibinfo {author} {\bibfnamefont {J.~N.}\ \bibnamefont {Moore}}, \bibinfo {author} {\bibfnamefont {T.}~\bibnamefont {Mano}}, \bibinfo {author} {\bibfnamefont {N.}~\bibnamefont {Shibata}},\ and\ \bibinfo {author} {\bibfnamefont {G.}~\bibnamefont {Yusa}},\ }\href {https://doi.org/10.1103/PhysRevResearch.4.L012040} {\bibfield  {journal} {\bibinfo  {journal} {Phys. Rev. Res.}\ }\textbf {\bibinfo {volume} {4}},\ \bibinfo {pages} {L012040} (\bibinfo {year} {2022})}\BibitemShut {NoStop}%
\bibitem [{\citenamefont {Kamiyama}\ \emph {et~al.}(2023)\citenamefont {Kamiyama}, \citenamefont {Matsuura}, \citenamefont {Moore}, \citenamefont {Mano}, \citenamefont {Shibata},\ and\ \citenamefont {Yusa}}]{kamiyamaAPL}%
  \BibitemOpen
  \bibfield  {author} {\bibinfo {author} {\bibfnamefont {A.}~\bibnamefont {Kamiyama}}, \bibinfo {author} {\bibfnamefont {M.}~\bibnamefont {Matsuura}}, \bibinfo {author} {\bibfnamefont {J.~N.}\ \bibnamefont {Moore}}, \bibinfo {author} {\bibfnamefont {T.}~\bibnamefont {Mano}}, \bibinfo {author} {\bibfnamefont {N.}~\bibnamefont {Shibata}},\ and\ \bibinfo {author} {\bibfnamefont {G.}~\bibnamefont {Yusa}},\ }\href@noop {} {\bibfield  {journal} {\bibinfo  {journal} {Appl. Phys. Lett.}\ }\textbf {\bibinfo {volume} {122}},\ \bibinfo {pages} {202103} (\bibinfo {year} {2023})}\BibitemShut {NoStop}%
\bibitem [{\citenamefont {W{\'o}js}\ \emph {et~al.}(2000)\citenamefont {W{\'o}js}, \citenamefont {Quinn},\ and\ \citenamefont {Hawrylak}}]{wojs2000charged}%
  \BibitemOpen
  \bibfield  {author} {\bibinfo {author} {\bibfnamefont {A.}~\bibnamefont {W{\'o}js}}, \bibinfo {author} {\bibfnamefont {J.~J.}\ \bibnamefont {Quinn}},\ and\ \bibinfo {author} {\bibfnamefont {P.}~\bibnamefont {Hawrylak}},\ }\href@noop {} {\bibfield  {journal} {\bibinfo  {journal} {Phys. Rev. B}\ }\textbf {\bibinfo {volume} {62}},\ \bibinfo {pages} {4630} (\bibinfo {year} {2000})}\BibitemShut {NoStop}%
\bibitem [{\citenamefont {Eytan}\ \emph {et~al.}(1998)\citenamefont {Eytan}, \citenamefont {Yayon}, \citenamefont {Rappaport}, \citenamefont {Shtrikman},\ and\ \citenamefont {Bar-Joseph}}]{eytan1998near}%
  \BibitemOpen
  \bibfield  {author} {\bibinfo {author} {\bibfnamefont {G.}~\bibnamefont {Eytan}}, \bibinfo {author} {\bibfnamefont {Y.}~\bibnamefont {Yayon}}, \bibinfo {author} {\bibfnamefont {M.}~\bibnamefont {Rappaport}}, \bibinfo {author} {\bibfnamefont {H.}~\bibnamefont {Shtrikman}},\ and\ \bibinfo {author} {\bibfnamefont {I.}~\bibnamefont {Bar-Joseph}},\ }\href@noop {} {\bibfield  {journal} {\bibinfo  {journal} {Phys. Rev. Lett.}\ }\textbf {\bibinfo {volume} {81}},\ \bibinfo {pages} {1666} (\bibinfo {year} {1998})}\BibitemShut {NoStop}%
\bibitem [{\citenamefont {Ilani}\ \emph {et~al.}(2004)\citenamefont {Ilani}, \citenamefont {Martin}, \citenamefont {Teitelbaum}, \citenamefont {Smet}, \citenamefont {Mahalu}, \citenamefont {Umansky},\ and\ \citenamefont {Yacoby}}]{ilani2004microscopic}%
  \BibitemOpen
  \bibfield  {author} {\bibinfo {author} {\bibfnamefont {S.}~\bibnamefont {Ilani}}, \bibinfo {author} {\bibfnamefont {J.}~\bibnamefont {Martin}}, \bibinfo {author} {\bibfnamefont {E.}~\bibnamefont {Teitelbaum}}, \bibinfo {author} {\bibfnamefont {J.}~\bibnamefont {Smet}}, \bibinfo {author} {\bibfnamefont {D.}~\bibnamefont {Mahalu}}, \bibinfo {author} {\bibfnamefont {V.}~\bibnamefont {Umansky}},\ and\ \bibinfo {author} {\bibfnamefont {A.}~\bibnamefont {Yacoby}},\ }\href@noop {} {\bibfield  {journal} {\bibinfo  {journal} {Nature}\ }\textbf {\bibinfo {volume} {427}},\ \bibinfo {pages} {328} (\bibinfo {year} {2004})}\BibitemShut {NoStop}%
\bibitem [{\citenamefont {Hayakawa}\ \emph {et~al.}(2013)\citenamefont {Hayakawa}, \citenamefont {Muraki},\ and\ \citenamefont {Yusa}}]{hayakawa2013real}%
  \BibitemOpen
  \bibfield  {author} {\bibinfo {author} {\bibfnamefont {J.}~\bibnamefont {Hayakawa}}, \bibinfo {author} {\bibfnamefont {K.}~\bibnamefont {Muraki}},\ and\ \bibinfo {author} {\bibfnamefont {G.}~\bibnamefont {Yusa}},\ }\href@noop {} {\bibfield  {journal} {\bibinfo  {journal} {Nat. Nanotechnol.}\ }\textbf {\bibinfo {volume} {8}},\ \bibinfo {pages} {31} (\bibinfo {year} {2013})}\BibitemShut {NoStop}%
\end{thebibliography}%

\appendix
\section{End Matter}
\textit{Appendix A: Excitation of the EMP by the Schottky gate.---}
Applying a voltage pulse to the excitation gate transiently modifies the local electrostatic potential near the mesa edge and thereby launches an EMP wave packet without injecting electrons from the gate into the 2DEG. The EMP wave packet propagates leftward along the mesa edge.

\textit{Appendix B: Applicability, coordinate convention, and asymptotic form of the Coulomb kernel.---}
Although Ref.~\cite{Aleiner} treats the classical Hall regime, this Coulomb kernel is not specific to that regime and also applies to the charged edge mode at $\nu=1/3$; the Hall-state dependence enters through the EMP velocity and transverse charge-density profile. Note that our $x$ and $y$ axes are defined oppositely to those in Ref.~\cite{Aleiner}, with edge propagation along $x$ in our case; we have relabeled their variables accordingly. The kernel behaves as $K_0(s)\approx -\ln s+\mathrm{const}$ for $s\ll1$ and decays as $K_0(s)\propto e^{-s}/\sqrt{s}$ for $s\gg1$. Such a transverse near field is therefore not expected to be unique to $\nu=1/3$, although whether it produces a similarly observable PL response in the integer or classical Hall regime remains to be investigated.

\textit{Appendix C: Delay-window coverage.---}
In Fig.~\ref{fig:y-time}(a), most of the $t>0$ region above the horizontal red feature is weakly red, whereas the corresponding $t<0$ region is nearly white. Although this faint red contrast becomes weaker toward the largest measured delay, it remains visible at the maximum measured delay of $t\sim 36$~ns. The displayed delay window covers only approximately $40$~ns of the $\sim 52$-ns repetition period. Because the measured response is periodic with the excitation, the peak-energy shift returns to the same pre-excitation value at the corresponding phase of the next cycle. Thus, the remaining return to the nearly white pre-excitation level occurs within the approximately $12.5$-ns portion of the cycle not shown in Fig.~\ref{fig:y-time}(a). In contrast, Fig.~\ref{fig:y-time}(b) includes a complete interval between successive horizontal features, and the peak-energy-shift contrast returns to the nearly white pre-excitation level before the next feature.

\textit{Appendix D: Comparison of the Line-B and Line-C PL spectra.---}
In the Line-B data measured with $T_\mathrm{rep} \sim 13$~ns, the spectra at $t = 10.0$~ns still show a pronounced voltage-pulse-induced modification relative to the spectra measured without the voltage pulse: the singlet feature remains enhanced and the triplet feature remains suppressed [Fig.~S2(a)]. In contrast, in the Line-C data measured with $T_\mathrm{rep} \sim 52$~ns, the spectra at $t = 9.9$~ns show a much weaker triplet suppression, as indicated by the recovered high-energy triplet feature [Fig.~S2(b)]. This comparison supports the interpretation that, under the 13-ns repetition condition, the triplet suppression does not fully relax before the next voltage pulse arrives.

\end{document}


\maketitle

\section{Voltage-Pulse Source and Biasing Conditions}

Voltage pulses were applied to the excitation gate, a Schottky gate capacitively coupled to the two-dimensional electron system (2DES). Each pulse transiently modifies the local electrostatic potential near the mesa boundary and launches an EMP wave packet without injecting electrons from the gate into the 2DES. Two voltage-pulse source configurations were used in this work, depending on the measurement condition. In both configurations, the applied waveform contained a rectangular voltage-pulse component with a duration of approximately $2$~ns.

In the first configuration, a rectangular voltage pulse was generated by a pulse generator and fed directly to the coaxial line connected to the excitation gate, without using a bias tee. 
This configuration was used for the measurements shown in Figs.~2, 3, 4(b), and 4(c) of the manuscript, and for the Line-B spectra in Fig.~\ref{fig:lineB_C_comparison}(a).
The pulse amplitude was set to $\pm 50$ or $\pm 100$~mV, as indicated in the corresponding figure. For the time-resolved measurements performed with this configuration, the repetition period was approximately $T_\mathrm{rep} = 13$~ns. The delay time coordinate used for these data is denoted by $t_1$.

In the second configuration, a negative-polarity rectangular voltage pulse was generated using an arbitrary waveform generator (AWG). The AWG output was switched from $+100$~mV to $-100$~mV for approximately $2$~ns and was combined with a $-0.1$~V dc bias using a bias tee before being applied to the excitation gate. This configuration was used for the measurements shown in Fig.~4(a) of the manuscript and the Line-C spectra in Fig.~\ref{fig:lineB_C_comparison}(b). For these measurements, a pulse picker was used to reduce the repetition rate of the optical pulses, extending the repetition period by a factor of four from approximately $T_\mathrm{rep} = 13$~ns to $T_\mathrm{rep} = 52$~ns. The electrical pulse train was synchronized to this repetition period. The delay time coordinate used for these data is denoted by $t_2$.

These two configurations are specified here to define the experimental conditions for each data set. In the manuscript, we discuss the relative temporal evolution within each delay time coordinate, rather than directly comparing the absolute delay time offsets between data sets acquired with different circuit configurations.

\begin{figure}[H]
  \centering
  \includegraphics[scale=1.0, pagebox=artbox, clip]{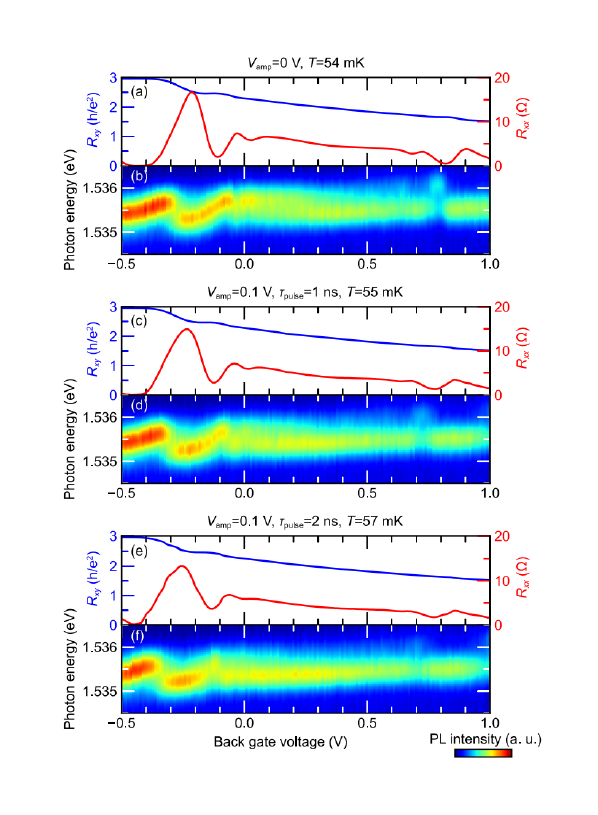}

  \caption{
    Representative simultaneous transport and PL measurements performed without and with nanosecond voltage-pulse excitation on a separate device. Panels (a), (c), and (e) show $R_{xy}$ (blue) and $R_{xx}$ (red) as functions of back-gate voltage, and panels (b), (d), and (f) show the corresponding PL-intensity maps. Panels (a) and (b) were measured without voltage-pulse excitation; panels (c) and (d) with $V_{\mathrm{amp}}=0.1$~V and $\tau_{\mathrm{pulse}}=1$~ns; and panels (e) and (f) with $V_{\mathrm{amp}}=0.1$~V and $\tau_{\mathrm{pulse}}=2$~ns. The measurement temperatures are indicated above the respective panels.
  }
  \label{fig:PL_transport}
\end{figure}

Figure~\ref{fig:PL_transport} shows representative simultaneous transport and PL measurements performed without and with nanosecond voltage-pulse excitation on a separate device. Although the transport traces show some evidence of pulse-induced heating, $R_{xx}$ still vanishes and $R_{xy}$ exhibits a plateau at $3h/e^2$, confirming that the 2DES remains in the $\nu=1/3$ state under pulsed excitation.

\section{Definition of Delay Time Origins}

As described in Sec. 1, two voltage-pulse source configurations were used in this work. Accordingly, we used two delay time coordinates, $t_1$ and $t_2$, with independently defined origins $t_{1,0}$ and $t_{2,0}$. The delay time is defined as the relative time delay between the optical pulse from the Ti:sapphire laser and the voltage pulse applied to the excitation gate. Since the measurement is performed in a stroboscopic manner, the delay time is a periodic coordinate determined by the repetition of the optical and electrical excitations. Therefore, the absolute origin of the delay time coordinate is arbitrary and does not have an independent physical meaning.

The coordinate $t_1$ is used for the data in Fig.~2, Fig.~3, Fig.~4(b), and Fig.~4(c), whereas $t_2$ is used for the data in Fig.~4(a).
Because these two coordinates have different origins, their absolute time offsets should not be directly compared.
However, the relative time evolution within each coordinate is meaningful.
Each coordinate should therefore be regarded as an operational time coordinate defined within the corresponding measurement condition.

We defined the origins $t_{1,0}$ and $t_{2,0}$ using the temporal response of the singlet PL amplitude $A_\mathrm{s}$ at the sample edge.
For each data set, the origin was chosen as the delay time at which the temporal gradient of $A_\mathrm{s}$ becomes most negative.
This convention corresponds to the steepest falling point of the PL response induced by the edge excitation and provides a consistent reference time within each delay time coordinate.

\begin{figure}[H]
  \centering
  \includegraphics[width=\linewidth]{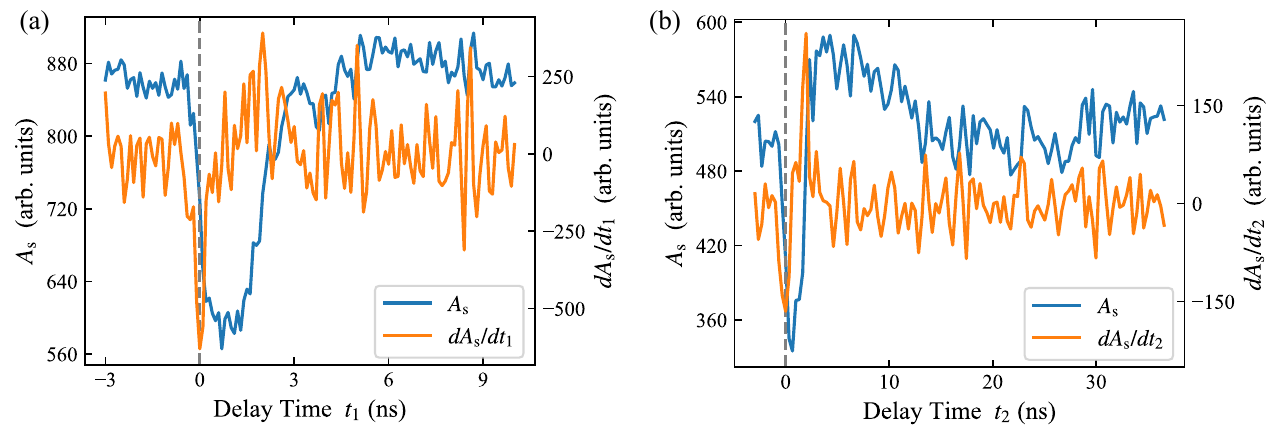}
  \caption{
    Definition of the time origins used in the delay time coordinates.
    (a) Time evolution of the singlet PL amplitude $A_\mathrm{s}$ measured at the sample edge on Line~B.
    The orange curve shows the temporal gradient $dA_\mathrm{s}/dt_1$, and the gray dashed line indicates the defined time origin $t_{1,0}$.
    (b) Time evolution of $A_\mathrm{s}$ measured at the sample edge on Line~C.
    The orange curve shows the temporal gradient $dA_\mathrm{s}/dt_2$, and the gray dashed line indicates the defined time origin $t_{2,0}$.
  }
  \label{fig:time-origin}
\end{figure}

For $t_1$, the origin $t_{1,0}$ was determined from the edge excitation measurement at the sample edge ($y = 0~\mu\mathrm{m}$) on Line~B (Fig.~1 of the manuscript), as shown in Fig.~\ref{fig:time-origin}(a).
For this measurement, a rectangular voltage pulse with $\pm 50~\mathrm{mV}$ amplitude and a pulse width of $2~\mathrm{ns}$ was generated by a pulse generator and applied to the excitation gate.
The gray dashed line indicates the defined origin $t_{1,0} = 0~\mathrm{ns}$.

For $t_2$, the origin $t_{2,0}$ was determined in the same manner from the PL response at the sample edge on Line~C (Fig.~1 of the manuscript), as shown in Fig.~\ref{fig:time-origin}(b).
For this measurement, a negative-polarity rectangular voltage pulse was generated by switching the AWG output from $+100~\mathrm{mV}$ to $-100~\mathrm{mV}$ for $2~\mathrm{ns}$ with a repetition period of approximately $52~\mathrm{ns}$.
The pulse train was offset with a $-0.1~\mathrm{V}$ dc bias using a bias tee and applied to the excitation gate.
The gray dashed line indicates the defined origin $t_{2,0} = 0~\mathrm{ns}$.

In the following sections, unless otherwise noted, the delay coordinate for each measurement configuration is denoted simply by $t$.

\section{PL Spectrum Fitting and Peak-Energy-Shift Analysis}

The collected photoluminescence (PL) was sent to a spectrometer equipped with a CCD detector.
The CCD output was recorded as a function of delay time $t$ between the laser and voltage pulses, position $(x, y)$, and photon energy $E$.
We denote the measured PL spectrum as $N(t, x, y, E)$.

For each fixed delay time and position, the PL spectrum as a function of $E$ was fitted by the sum of two Lorentzian functions and a constant background:
\begin{equation}
    f(E) =
    \frac{A_\mathrm{s} \gamma^2}{(E - E_\mathrm{s})^2 + \gamma^2}
    + \frac{A_\mathrm{t} \gamma^2}{(E - E_\mathrm{t})^2 + \gamma^2}
    + C .
    \label{eq:spectrum_fitting}
\end{equation}
The parameters $E_\mathrm{s}$ and $E_\mathrm{t}$ are the peak energies, and $A_\mathrm{s}$ and $A_\mathrm{t}$ are the peak amplitudes.
Because of the normalization used in Eq.~\eqref{eq:spectrum_fitting}, $A_\mathrm{s}$ and $A_\mathrm{t}$ correspond to the peak heights, not to the integrated PL intensities.
The parameter $2\gamma$ is the full width at half maximum (FWHM), and $C$ represents a constant background of the CCD output.

The linewidth parameter $\gamma$ was taken to be common to the singlet and triplet components in each fit to reduce the number of free fitting parameters and to stabilize the extraction of the peak energies.
The peak energies extracted using Eq.~(\ref{eq:spectrum_fitting}) were used to construct the PL peak-energy-shift maps shown in Fig.~4. At each measurement position, the reference peak energy $E_\mathrm{t}^{0}(x,y)$ for Fig.~4(a) was defined as the average of $E_\mathrm{t}(t,x,y)$ evaluated at $t=-2.965$, $-2.64$, and $-2.31$~ns. For Figs.~4(b) and 4(c), $E_\mathrm{t}^{0}(x,y)$ was defined as the average evaluated at $t=-3$, $-2$, and $-1$~ns in the first period, and the same reference value was used for both the first and second periods. The triplet peak-energy shift was then calculated as
\begin{equation}
\Delta E_\mathrm{t}(t,x,y)=E_\mathrm{t}(t,x,y)-E_\mathrm{t}^{0}(x,y).
\label{eq:triplet_energy_shift}
\end{equation}

\section{Comparison of PL Spectra at $t \simeq 10$~ns along Lines B and C}

\begin{figure}[H]
  \centering
  \includegraphics[width=0.6\linewidth]{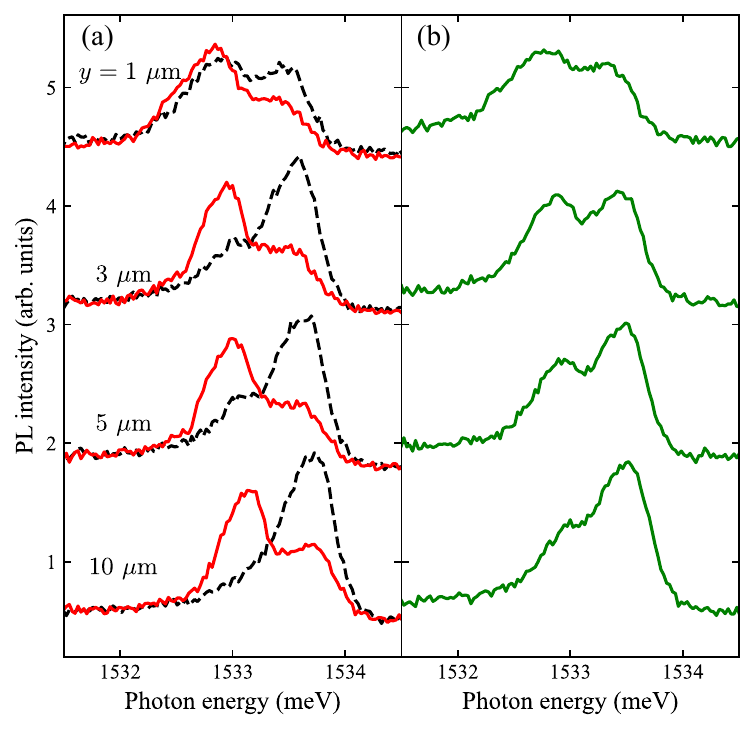}
  \caption{
    Comparison of PL spectra at $t \simeq 10$~ns along Lines B and C.
    (a) PL spectra extracted from the Line-B time-resolved data set used for Fig.~4(b) of the manuscript at $t = 10.0$~ns, measured with $T_\mathrm{rep} \sim 13$~ns. Red solid curves show spectra measured with the voltage pulse, and black dashed curves show spectra measured without the voltage pulse.
    (b) PL spectra extracted from the Line-C time-resolved data set used for Fig.~4(a) of the manuscript at $t = 9.9$~ns, measured with $T_\mathrm{rep} \sim 52$~ns. Green solid curves show spectra measured with the voltage pulse.
    In both panels, spectra are shown at $y = 1$, $3$, $5$, and $10~\mu\mathrm{m}$ from top to bottom and are vertically offset for clarity.
  }
  \label{fig:lineB_C_comparison}
\end{figure}

Figure~\ref{fig:lineB_C_comparison} compares representative PL spectra at $y = 1$, $3$, $5$, and $10~\mu\mathrm{m}$. The Line-B spectra in Fig.~\ref{fig:lineB_C_comparison}(a) were extracted from the time-resolved data set used for Fig.~4(b) of the manuscript at $t = 10.0$~ns, with $T_\mathrm{rep} \sim 13$~ns. The Line-C spectra in Fig.~\ref{fig:lineB_C_comparison}(b) were extracted from the data set used for Fig.~4(a) at $t = 9.9$~ns, with $T_\mathrm{rep} \sim 52$~ns.

In Fig.~\ref{fig:lineB_C_comparison}(a), the Line-B spectra still show a pronounced voltage-pulse-induced modification relative to the spectra measured without the voltage pulse: the singlet feature is enhanced and the triplet feature is suppressed. In contrast, the Line-C spectra in Fig.~\ref{fig:lineB_C_comparison}(b) show much weaker triplet suppression, as indicated by the recovered triplet feature on the higher-energy side.

